\documentclass[Afour,sageh,times]{sagej}
\setcitestyle{numbers} 

\usepackage{moreverb,url}

\usepackage[colorlinks,bookmarksopen,bookmarksnumbered,citecolor=red,urlcolor=red]{hyperref}

\newcommand\BibTeX{{\rmfamily B\kern-.05em \textsc{i\kern-.025em b}\kern-.08em
T\kern-.1667em\lower.7ex\hbox{E}\kern-.125emX}}

\usepackage{array}
\newcommand{\PreserveBackslash}[1]{\let\temp=\\#1\let\\=\temp}
\newcolumntype{C}[1]{>{\PreserveBackslash\centering}p{#1}}
\usepackage{amssymb}
\usepackage{amsmath}
\usepackage{pgfgantt}
\usepackage{pgfplots}
\usepackage{natbib}
\theoremstyle{definition}
\newtheorem{definition}{Definition}[section]

\begin{document}

\runninghead{J. Figueira}

\title{A Survey on Semantic Steganography Systems}

\author{Jo\~{a}o Figueira\affilnum{1}}
%\author{Anonymous for review}
\affiliation{\affilnum{1}Instituto Superior T\'{e}cnico, Portugal}

\email{joaoperfig@gmail.com}

\begin{abstract}
Steganography is the practice of concealing a message within some other carrier or cover message. It is used to allow the sending of hidden information through communication channels where third parties would only be aware of the explicit information in the carrier message. With the growth of internet surveillance and the increased need for secret communication, steganography systems continue to find new applications. In semantic steganography, the redundancies in the semantics of a language are used to send a text steganographic message.
    In this article we go over the concepts behind semantic steganography and propose a hierarchy for classifying systems within the context of text steganography and steganography in general. After laying this groundwork we list systems for semantic steganography that have been published in the past and review their properties.
    Finally, we comment on and briefly compare the described systems.
\end{abstract}


\maketitle

\section{Introduction and Background}

Steganography systems describe methods for taking an "innocuous" message, called \textit{covertext} and embed it with some \textit{plaintext} message that is desired to remain hidden, outputting a \textit{stegotext}. This \textit{stegotext} is a slightly altered version of the \textit{covertext} that is still "innocuous" and from which the \textit{plaintext} is extractable. Effectively, steganography is the process of encrypting a message and having the output appear to be a non-encrypted message.

In certain contexts, a communication channel provider might refuse to relay messages that it can see are encrypted and does not know are trustworthy. More recently, some governments have been planning to outlaw or to regulate the usage of some encryption systems and others have already begun to do so. Steganography finds applications in these situations and will continue to do so with the growing threat of mass surveillance.

Semantic steganography is the branch of text steganography that uses redundancies in the vocabulary of natural languages as the space for the \textit{plaintext} message~\cite{blue}.

\subsection{The Steganographic Process}

\begin{figure}[t]

\includegraphics[width=0.48\textwidth]{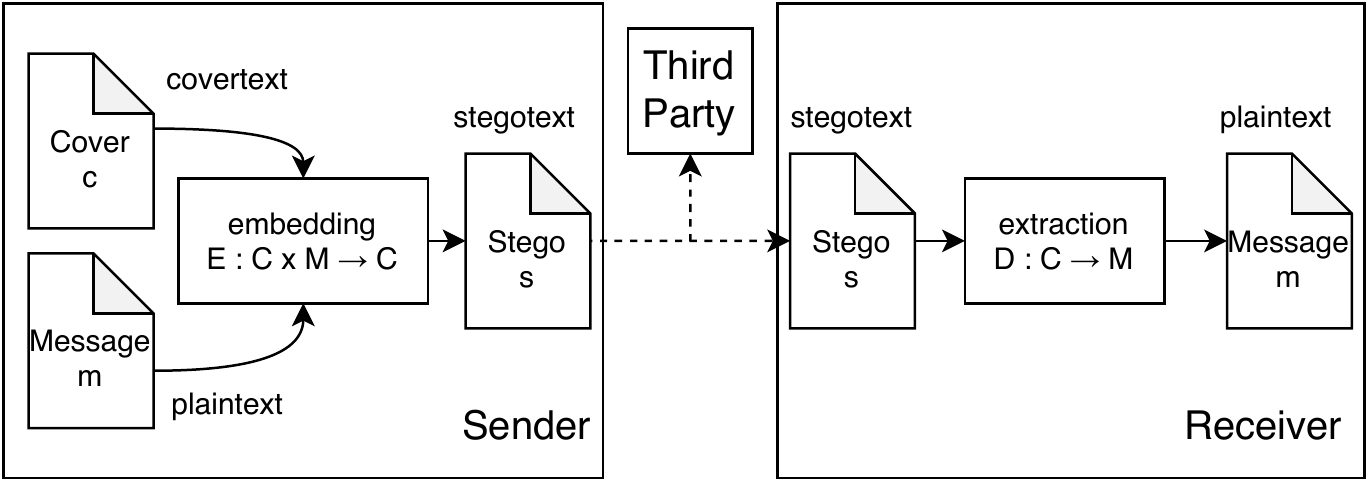}

\caption{Diagram showing a generic setting for steganography. The sender embeds the \textit{plaintext} message $m$ into the \textit{covertext} $c$, producing the \textit{stegotext} $s$. This \textit{stegotext} is sent to the receiver who uses the extraction function to recover the original \textit{plaintext}. A third party might see the \textit{stegotext} but it should be innocuous enough that no suspicion would be raised to the fact that it is carrying a message.}
\label{stegoprocess}
\end{figure}

As described by Kingslin in~\cite{blue}, a steganography system can be divided into two components:
\begin{itemize}
  \item An embedding or injection method, where a \textit{covertext} is modified to receive the \textit{plaintext}, outputting the \textit{stegotext}. These functions make use of the redundancies in the \textit{covertext} and exploit them as the space in which the \textit{plaintext} will be inserted. This is performed by the message sender. 
  \item An extraction method, where the \textit{plaintext} is extracted from the \textit{stegotext} (in some systems, the original \textit{covertext} can also be obtained here). This is done by the message receiver.
\end{itemize}

A diagram explaining the usage of these two functions to hide and send messages can be seen in Figure~\ref{stegoprocess}.

\theoremstyle{definition}
\begin{definition} \label{formalpurestego}
A steganography system (or scheme) can be defined as a quadruple $<C, M, E, D>$, where $C$ is the set of possible \textit{covertexts} (messages that are innocuous and would not raise suspicion to a third party), $M$ is the set of possible \textit{plaintexts} (the set of all messages over the \textit{plaintext} alphabet $\Sigma$, $\Sigma ^*$), with $|C| \geq |M|$, $E : C \times M \rightarrow C$ is the embedding or insertion function, $D : C \rightarrow M$ is the extraction function. The property $D(E(m)) = m$ is verified for all $m \in M$.  The embedding and extraction functions might take additional parameters, depending on the system.

\end{definition}

\subsection{A Hierarchy for Text Steganography Systems}

Most Steganography methods can embed a hidden message of any nature into a cover message of a specific nature, \textit{i.e.} the embedding and extraction functions will be constructed for the specific given source of covers \cite{stegobook}. As such, Steganography methods are usually classified according to the type of cover message they work with \cite{monika,blue,types1,types2}. In the digital age, with such a variety of media types and file formats, steganography methods have been developed for almost all types of possible cover messages. The main ones include images, audio, and text (for a video cover message, image and audio steganography systems can be applied independently).

Text steganography is the family of steganographic systems that use text as the cover message. Historically, writing has been one of the oldest forms of communication over long distances. As such, it is likely that text steganography is the oldest form of steganography. Despite this, text steganography is still seen as the most difficult kind of steganography. As Sharma describes it in \cite{survey}, this is because a text file lacks a large scale redundancy of information in comparison to the other digital media formats.

\begin{figure}[t]

\includegraphics[width=0.48\textwidth]{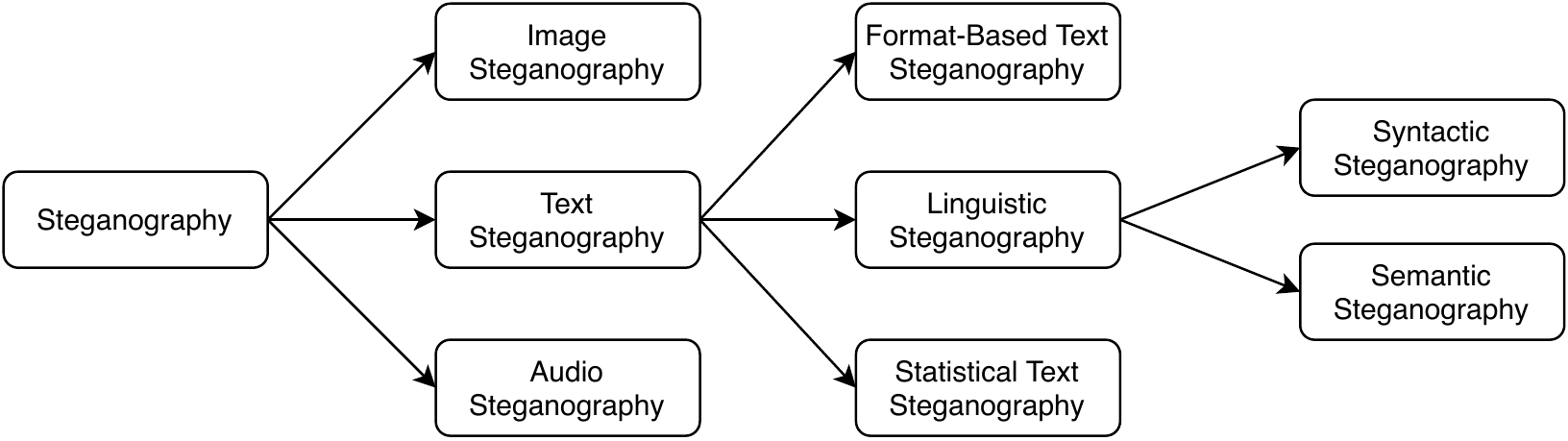}

\caption{Diagram showing the proposed hierarchy of the major families of steganographic systems.}
\label{stegotypes}
\end{figure}

Due to its longer history, text steganography is an area of research that has seen the development of some very different approaches, these differ mostly in what elements of the message are affected in order to receive the hidden message. In this section we provide and describe a hierarchy for the classification of text steganography systems. A diagram of this hierarchy can be seen in Figure~\ref{stegotypes}.

\subsubsection{Format-Based Text Steganography}
Text steganography systems that alter the formatting of the text are called format-based steganography systems. Altering the formatting of the text might involve things such as slightly altering the size or color of letters, moving words or sentences a few millimetres, or even adding extra spaces between words~\cite{bender,monika,blue,types1,types2}. These systems are the most commonly used for text steganography.

In \cite{bender}, Bender states that these systems can be further divided into two categories: "Soft-copy safe systems", which are the systems in which the hidden message is not lost if the text is copied onto a different file, these include the insertion of spaces between words; And "Hard-copy safe systems" which are systems in which the text formatting is closely related to the specific file format of the text, in these systems the hidden message is likely to be lost if the text is copied onto some other file, Bender \cite{bender} described that these systems can be treated as a "highly structured image".

These systems have the problem that many communication channels, such as online messaging systems, will detect and fix what they  consider "formatting errors" and the hidden message can be easily lost. Another vulnerability, described by Agarwal \cite{monika}, is that these changes in formatting can be easily detected by opening the text in a word processor.

\subsubsection{Statistical Text Steganography}
Statistical text steganography, often also called random text steganography \cite{inf_hiding2,monika,blue}, is the branch of text steganography that deals with hiding information in statistical properties of the \textit{covertexts}. To achieve this, most statistical steganographic systems usually deal with generating the \textit{stegotext} itself (a process mentioned in Section~\ref{coversynthesis}). The \textit{stegotext} is generated in such a way that the desired statistical properties of the text are verified.

The most simple example of such a system would be the "cover lookup" system described by Kaufmann in \cite{stegobook}. In this system, the message sender has a set of possible cover messages (or generates them) and simply selects the one that, using a specific hashing function, hashes to a desired \textit{plaintext}. More advanced systems of this branch might, for example, hide information in the frequencies of certain words or letters.

\subsubsection{Linguistic Text Steganography}
Text steganography systems that deal with the linguistic properties of the \textit{covertext} are called linguistic steganographic systems \cite{monika,blue,types1}. These systems perform alterations on the text itself and exploit the ambiguities or redundancies of natural languages.

As described by Kinglslin~\cite{blue} and Singh~\cite{types1}, the family of linguistic steganography systems can be further divided according to which linguistic properties of the text are being used to embed the \textit{plaintext}. As such, the following two sub-families of linguistic steganography can be formalized:

\begin{itemize}

\item \textbf{Syntactic Text Steganography}
Linguistic steganography systems that deal with the syntax of text are called syntactic text steganography systems. Such systems might change the grammatical structures of sentences to embed a hidden message. Simpler systems in this family might simply add or remove commas from text in places where their necessity is arguable (such as the Oxford comma).

\item \textbf{Semantic Text Steganography}
Semantic text stenography is the branch of text steganography that uses the redundancy of words as the space for the hidden message. Steganographic systems in this family replace words in a cover message with their synonyms.

Trivial implementations of such systems label words and their synonyms with a binary value. The message sender identifies the words that can be replaced in the covertext, and, depending on the desired bit from the \textit{plaintext}, will choose to keep the original word or replace it with its synonym. The message receiver will do the same process and identify the message sender's choices to determine the hidden message bits.

\end{itemize}

\subsection{Classifications for Embedding Functions}
The embedding and extraction functions are the defining element of a steganographic system. As inverse functions, these two methods are co-dependant and need to be jointly defined. For their relevance, steganographic systems can be classified according to the working principles of these functions. The following classifications where proposed by Kaufmann in \cite{stegobook}:

\begin{itemize}

\item\textbf{Steganography by Cover Modification} Steganography systems in which the embedding function alters an existing \textit{covertext} are called steganography by cover modification. This is the most common working principle of steganographic systems and is the one shown in Figure~\ref{stegoprocess}. In \cite{alphabet}, Osman considers that this category can be further divided into substitution-based systems, in which parts of the cover message are replaced; and injection-based systems, in which new elements are inserted into the cover message.

\item\textbf{Steganography by Cover Synthesis}\label{coversynthesis}
The generation of a \textit{stegotext} based on the \textit{plaintext} is called steganography by cover synthesis (or generation). This type of steganography can be seen as difficult as it might be hard to generate a cover message that is natural and innocuous.

\item\textbf{Steganography by Cover Lookup}
Steganography by cover lookup describes steganographic systems in which the cover messages are preexisting and not modified in any way. In these systems, the message sender will use the extraction function on all available cover messages and choose the one that produces the desired \textit{plaintext}.

\end{itemize}

\subsection{On The Security of Semantic Steganography Systems}\label{security}
The primary objective of any steganography system is to provide a hidden channel for communication, such that third parties can intercept the cover messages and not be suspicious that these messages are carrying a hidden embedded message. Some third parties, might, however, be aware of the possibility of usage of steganography in a certain communication channel. In this situation, they might use the extraction functions of some steganography systems to "screen" messages for possible hidden embedded messages. Because of this, it is always ideal to first encrypt the hidden message using, for example, some simple symmetric-key encryption algorithm. If the extraction function can be used on any cover message and have some output, the natural randomness of some \textit{covertext} should be indistinguishable from the \textit{ciphertext} produced by some cryptosystem \cite{ladybook}. This means that not only is the message encrypted, but there is also no evidence that any steganography system was used, as any message would give a seemingly random output from the extraction function.

\section{Semantic Steganography Systems}
The following is our survey of existing systems for semantic steganography that have been published in the past.

\subsection{Synonym Table Steganography Systems}
Semantic steganography systems use the redundancy in the words of natural languages as the space for a hidden message. The most trivial implementation of such a system would be one that replaces words in the \textit{covertext} with their synonyms. In our survey, the majority of such systems make usage of a synonym table (exemplified in Table~\ref{synonymtable}) that is shared between the message sender an receiver. These tables, of usually two columns, pair words with their synonyms.

\begin{table}
\caption{Example of a short synonym table, used by Bender in \cite{bender}.\label{synonymtable}}
\centering
\begin{tabular}{|C{4cm}|C{4cm}|}
\hline
big & large\\
small & little\\
chilly & cool\\
smart & clever\\
spaced & stretched\\
\hline
\end{tabular}
\end{table}

In these systems, the hidden message is encoded into the choice of synonyms that was used in the \textit{covertext}. This way, each word in the \textit{covertext} (that can be replaced by a synonym) will encode a character of the \textit{plaintext}, corresponding to which column of the synonym table it is in.

In the approaches described by Bender~\cite{bender}, Rafat~\cite{smskey}, and Shirali-Shahreza \cite{sms,britishenglish}, the \textit{plaintext} is first converted into a binary string. This way, a two-column synonym table can be used to encode the hidden message (there is one column for each character of the hidden message alphabet $\Sigma = \{0, 1\}$).

\subsubsection{Embedding Method}
\begin{figure}[t]
\centering
\includegraphics[width=0.48\textwidth]{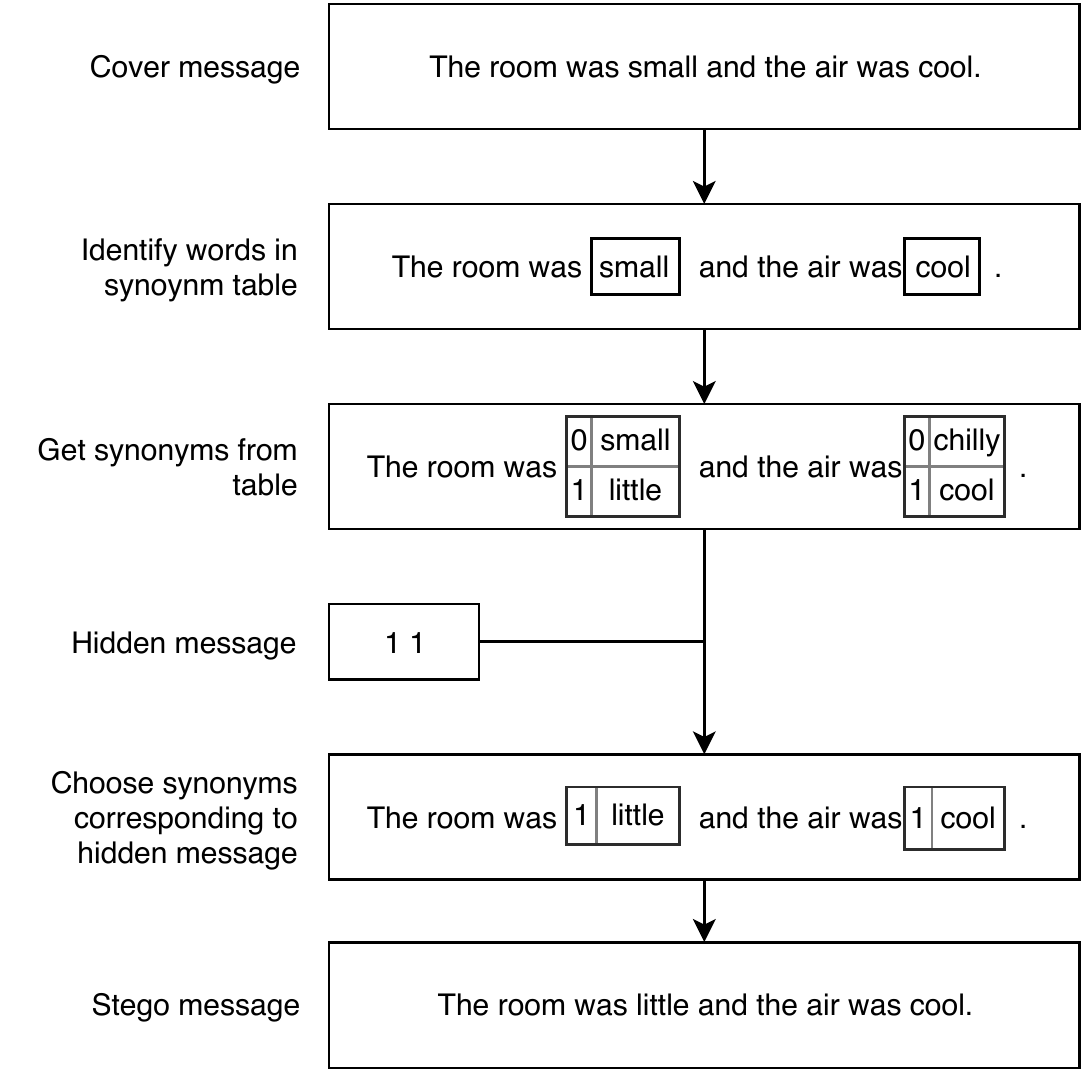}
\caption{Diagram exemplifying the embedding process of the \textit{plaintext} "11" into the \textit{covertext} "The room was small and the air was cool." using a semantic steganographic system with the synonym table shown in Table~\ref{synonymtable}.}
\label{synembed}
\end{figure}

In all of these systems \cite{bender,sms,smskey,britishenglish}, the embedding method functions as follows, for a given \textit{covertext} and \textit{plaintext}:
\begin{enumerate}
    \item The \textit{plaintext} is converted into an alphabet $\Sigma$ such that $|\Sigma| = c$, where $c$ is the number of columns in the synonym table.
    \item The \textit{covertext} is scanned and occurrences of words in the synonym table are identified.
    \item The $n^{th}$ identified word of the \textit{covertext} is replaced with a synonym from the table's column corresponding to the $n^{th}$ character of the \textit{plaintext}.
\end{enumerate}{}
This embedding method is further clarified in Figure~\ref{synembed}.

\subsubsection{Extraction Method}
\begin{figure}[t]
\centering
\includegraphics[width=0.48\textwidth]{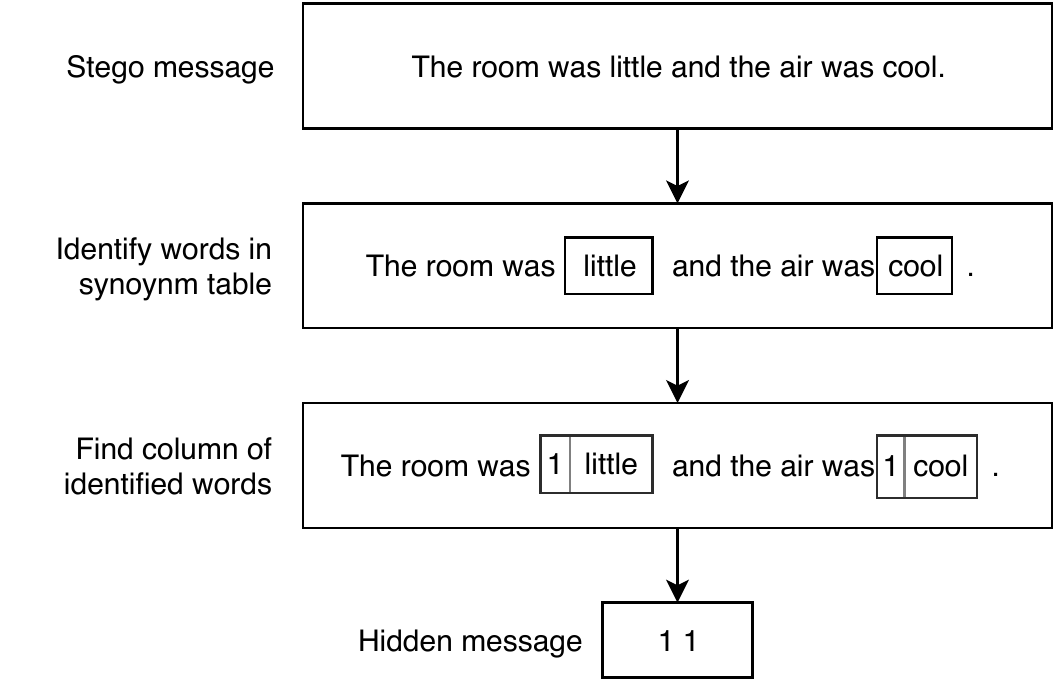}
\caption{Diagram showcasing the extraction process from the \textit{covertext} generated in figure \ref{synembed}. This process requires the same synonym table shown in Table~\ref{synonymtable}.}
\label{synextract}
\end{figure}

The \textit{stegotext} generated by the message sender using the aforementioned embedding method is sent to the message receiver which will apply the corresponding extraction method. The extraction method for these systems can be described as follows:
\begin{enumerate}
    \item The \textit{stegotext} is scanned and occurrences of words in the synonym table are identified.
    \item The $n^{th}$ character of the \textit{plaintext} will correspond to the column of the $n^{th}$ identified word of the \textit{stegotext}.
\end{enumerate}{}
This extraction method is further clarified in Figure~\ref{synextract}.

\subsubsection{Synonym Tables} The authors in \cite{bender,survey} explain the usage of these systems in a generic context. In \cite{lex}, the author described this as the "naive" implementation of a semantic steganographic system. In regard to the synonym table itself, these authors simply described it as a table that pairs words with interchangeable synonyms and offered no source as to how the table would be constructed or to which specific words could be used. It is not entirely trivial how these tables should be constructed. Words that seem synonymous in certain contexts might not be interchangeable in other contexts \cite{bender,lex}. 

In \cite{britishenglish}, Shirali-Shahreza explored the usage of words that have different spellings in American English and European English (for example, "Candy" and "Sweets"). This approach is advantageous in that there should be no occurrence of two words in the table not always being interchangeable. 

Acronyms and their unabbreviated counterparts can also be seen as synonyms. In \cite{sms}, the authors explored the application of the abbreviations and acronyms commonly used in SMS messages for such a system. Their approach would ideally be applied to SMS messaging where the usage of the aforementioned acronyms would be most innocuous (such as using "NP" instead of "No problem"). An obvious disadvantage of this, however, is that SMS messaging is not an ideal channel for steganography, due to the character limit of text messages. In the described system, only about three bits of information could be sent per message. As such, hundreds of messages would be needed to send a single paragraph of \textit{plaintext}, which would not be very innocuous.

The work of Shirali-Shahreza \cite{sms} was extended by Rafat's research \cite{smskey}. In this article, the author explored the expansion of the security of the system by using a \textit{stego-key} to shuffle elements of the synonym table between the left and right columns. For his implementation, the author used a process of XoR-Encryption supported by a Linear Feedback shift register to perform the shuffling. This system is more secure in that a third party that might know the system would still be unable to extract the hidden information without the \textit{stego-key}. 

Rafat's approach does, however, have some vulnerabilities. Shuffling the synonym table for security means that expressions in the table will always correspond to the same character of the hidden alphabet. For example, a third party that knows the system but does not know the \textit{stego-key} would not know which bit is encoded in an instance of the acronym "np" in the \textit{covertext}. However, this third party will know that all instances of "np" encode the same bit value which consequentially is different from the bit value of all instances of "no problem". This means the system might be vulnerable to some statistical analysis methods. A simpler and safer approach to security would be the one described in section~\ref{security}.

\subsection{Variable Synonym Cardinality Steganography Systems}

In the examples described in previous section, the synonym table has a set number of columns and, as such, all words in such a synonym table are restricted to having that set number of possible replacements (usually just one, for the embedding of a binary string). This is rather restrictive since some words (usually the most common ones) can have a large number of synonyms. These words have a potential to encode more information that is not being exploited by the described system.

\begin{figure}[t]
\centering
\includegraphics[width=0.48\textwidth]{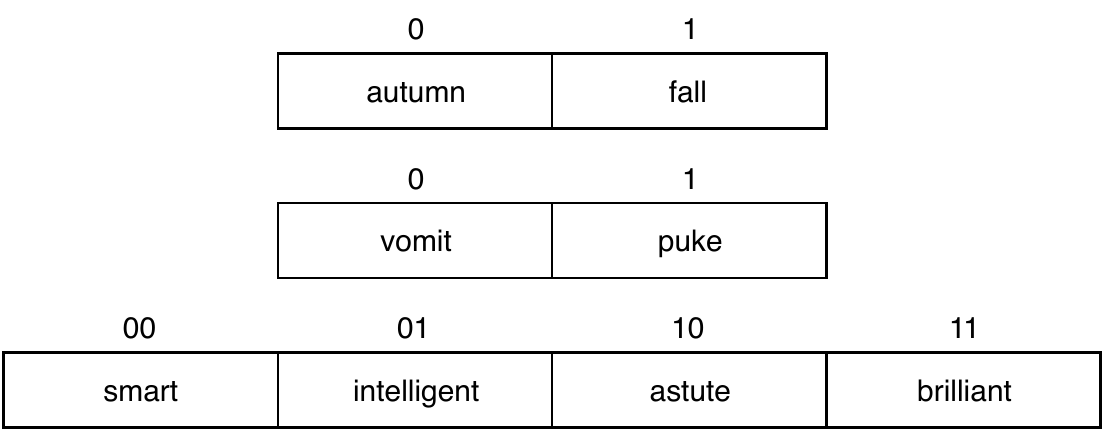}
\caption{Examples for sets of synonyms and the bits they can encode, as described by Winstein in \cite{lex}.}
\label{multisyn}
\end{figure}

The most trivial solution for this problem is the one described by Winstein~\cite{lex} in his description for a "naive algorithm". The described approach groups words into sets of mutually interchangeable synonyms. The system embeds a binary message into the \textit{covertext}, each word can embed as many bits as the base two logarithm of the number of words in its synonym set. As such, the number of elements in these sets of synonyms is restricted to being some power of 2. This approach is exemplified in Figure~\ref{multisyn}. A similar approach is also used in \cite{large_scale} and \cite{raresyn}.

\subsection{Winstein's Ideal Coding}\label{idealc}

\begin{figure}[t]
\centering
\includegraphics[width=0.48\textwidth]{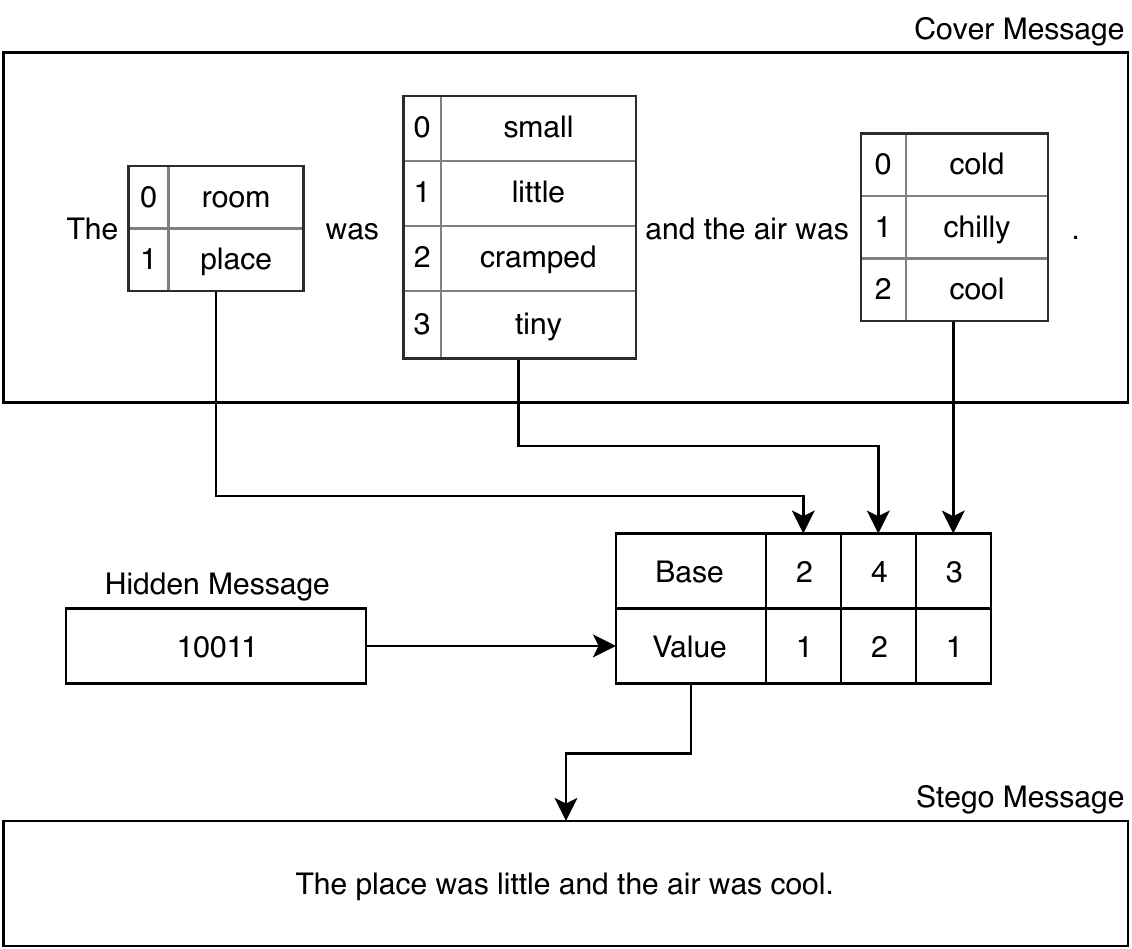}
\caption{Diagram showcasing the embedding of the hidden message "10010" into the \textit{covertext} "The room was small and the air was cool.", using the \textit{multi-base number} approach described by Winstein in \cite{lex}.}
\label{lexdiag}
\end{figure}

In \cite{lex}, Winstein improves on the aforementioned "naive algorithm" by proposing a related system in which the synonym sets can have any number of words (as opposed to only powers of 2). His proposal consists on converting the hidden message into a \textit{multi-base number} (each digit may have a different base), where each digit corresponds to a word in the synonym table, and the base of each digit is the number of replacements that word can have.  This solution can be visualised with the diagram in Figure~\ref{lexdiag}.

\subsection{Mimic Functions}\label{mimicsection}
A well known approach for semantic steganography is the one proposed by Wayner in his articles \cite{mimic1} and \cite{mimic2}. Here Wayner described the construction of mimic functions and their applications for text steganography.

A mimic function $f$ is described as the function that alters the statistical properties of a text file $A$ to be the same as some other file $B$. Formally, if $p(t, A)$ is the probability of a substring $t$ occurring in $A$, then the mimic function $f$ encodes $A$ so that $p(t, f(A))$ approximates $p(t, B)$.

Wayner introduces mimic functions as the inverse of Huffman compression functions. A Huffman function (or Huffman code) is a type of optimal prefix code that is commonly used for lossless data compression, it was first proposed by Huffman \cite{huffman}. 

The proposed approach is to construct the Huffman compression functions for the files $A$ and $B$, $f_A$ and $f_B$. The inverse of $f_B$, $g_B$ is then computed. The composite function $g_B(f_A(A))$ is the first order mimic function that converts $A$ to have the statistical properties of $B$. Larger order mimic functions can be computed by joining sequences of $n$ characters together (for an $n^{th}$ order mimic function) and interpreting them as being a single character.

This system can be seen as a cover synthesis steganographic system (the cover message is generated for the specific \textit{plaintext}) in which, for a hidden message $A$, and a cover message $B$, the mimic function $g_B(f_A(A))$ will generate a \textit{stegotext} message that has the statistical properties of the \textit{covertext} $B$. The \textit{stegotext} outputted by this system will be text that contains word or even short expressions found in the \textit{covertext} but that lacks any grammatical structure or sense. For a human third party, this \textit{stegotext} will obviously raise suspicion. Wayner improved on his system by joining it with context-free grammars to ensure the sentences maintain grammatical consistency. This improved the iniquity of the \textit{stegotext}, but it still remained mostly devoid of meaning.

\subsection{Markov Chain Based Text Steganography}
In \cite{daikov1}, Dai introduced the usage of Markov chains for text staganography, this research was continued in \cite{daikov2}. Dai's proposal involves constructing a Markov model for the desired \textit{covertext}.

A Markov model constructed from some text corpus would maintain the probabilities of any two consecutive words appearing in the corpus $p(w_i | w_{i-1}) =  {count(w_{i-1}, w_i)}/{count(w_{i-1})}$. The model could be used to construct new text samples that mimicked the statistical properties of the corpus (much like the mimic functions described in Section~\ref{mimicsection}). Higher order Markov models can be used, these models take in account more words to provide a more accurate probability of the next. An $n^{th}$ order Markov model would compute and use the probabilities $p(w_i | w_{i-1}, w_{i-2}, ..., w_{i-n})$.

\begin{figure}[t]
\centering
\includegraphics[width=0.48\textwidth]{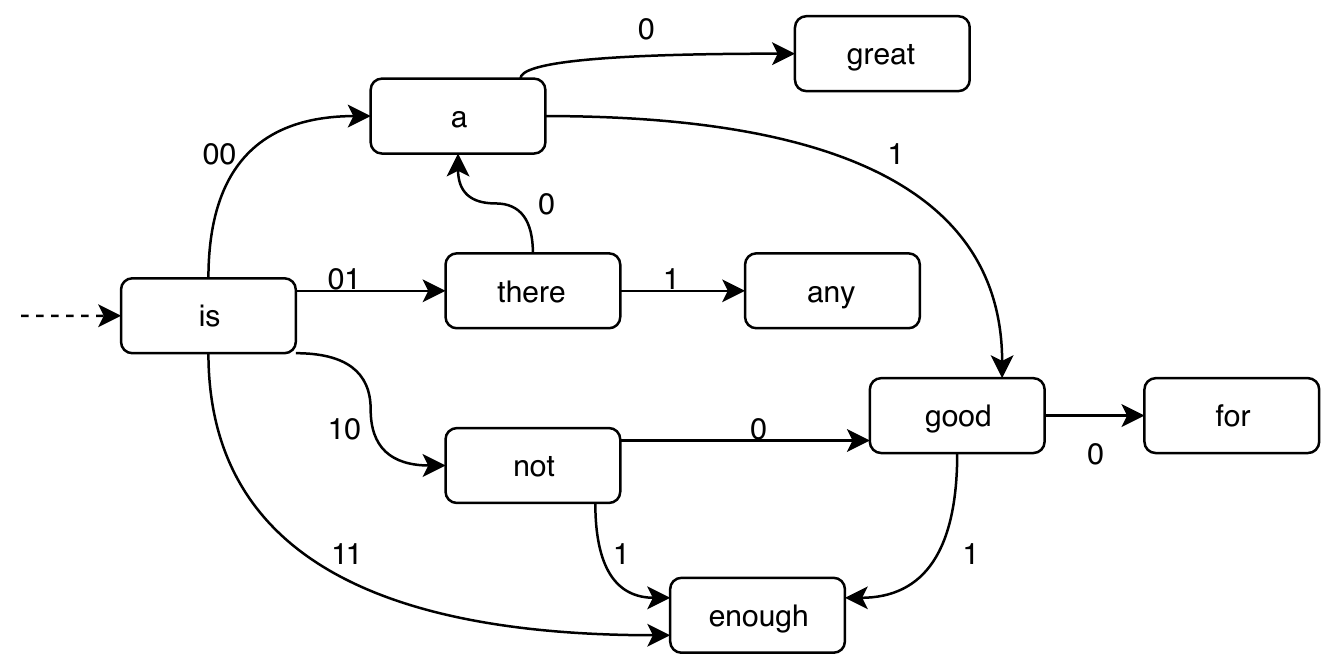}
\caption{Diagram of a steganography system constructed using a Markov model, as described in \cite{daikov1}. To exemplify the function of the model, the hidden message "0100" would synthesize the \textit{stegotext} "is there a great", while "1001" would synthesize the \textit{stegotext} "is not good enough".}
\label{markovdiag}
\end{figure}

In Dai's approach, the transitions of the Markov model are labelled with parts of the hidden message. To synthesise the \textit{stegotext}, it is only necessary to use the \textit{plaintext} to determine the sequence of state transitions that is done on the model. This process is exemplified Figure~\ref{markovdiag}.

\subsection{Moraldo's Fixed Size Steganography}
In \cite{markov}, Moraldo described how Dai's Markov systems produce "unnatural" looking text by not taking into account the probability of transitions. With the way that transitions are labelled, any outgoing transition from any given state has the same probability of occurring in a \textit{stegotext}.

\begin{figure}[t]
\centering
\includegraphics[width=0.48\textwidth]{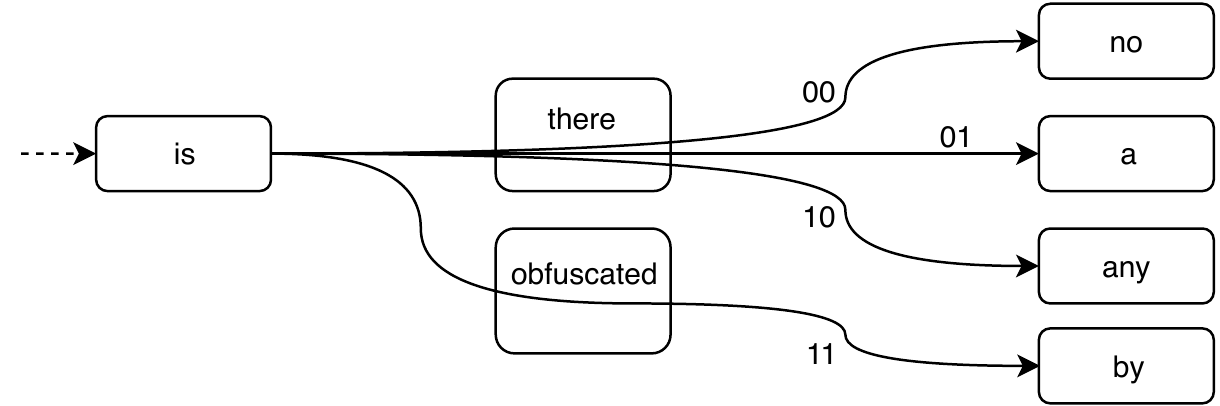}
\caption{Diagram of grouped state transitions as described by Moraldo \cite{markov}. By grouping two consecutive transitions, the transition "is there" is made more probable than "is obfuscated", as expected in text. In Dai's approach, these transitions would be equiprobable.}
\label{markov2diag}
\end{figure}

Moraldo's solution involves grouping multiple consecutive transitions together and labeling these groups with parts of the hidden message. More probable state transitions will occur in more of these labelled groups. This way, the resulting \textit{stegotext} will have more natural word sequences that occur with the frequency that is expected of a real text. This system is exemplified in Figure~\ref{markov2diag}.

\subsection{Markov Chain and Huffman Coding Based Text Steganography}\label{markuffsec}

\begin{figure}[t]
\centering
\includegraphics[width=0.48\textwidth]{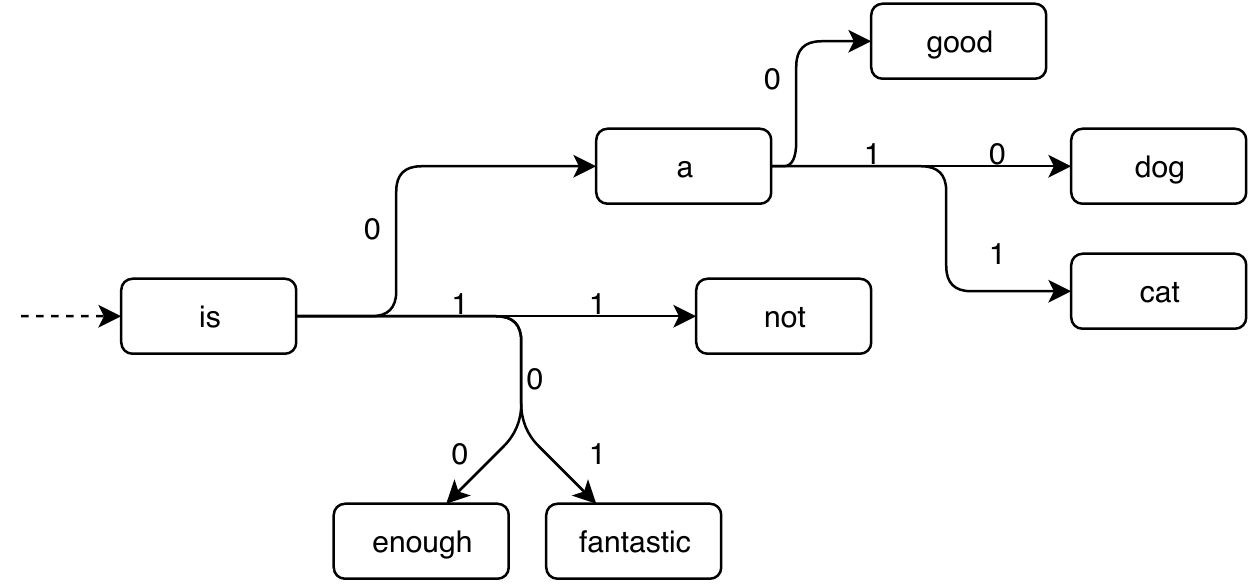}
\caption{Diagram exemplifying the usage of "chained" Huffman trees to label transitions in a Markov model, as described by Zhongliang \cite{markuff}.}
\label{markuffdiag}
\end{figure}

In \cite{markuff}, the authors also explore the problem of ensuring a natural probability distribution of transitions on a Markov based steganographic system. For their approach, the authors make use of Huffman coding to construct a tree for the transitions at each step of the Markov model. More frequent transitions are labelled with shorter labels and are thus more likely to appear in the hidden message. This is exemplified in Figure~\ref{markuffdiag}. This system shares a lot of similarities with the mimic functions described by Wayner~\cite{mimic1} and with Moraldo's approach~\cite{markov}.

\section{Conclusion}

Text steganography systems continue to be tools with very niche and circumstantial applications but that are sure to see increased usage with the growing trend of online surveillance and censorship. We found that the literature on the topic was sparse and and not coherent across publications and hope that our survey helps standardize approaches to semantic steganography.

In regards to the listed approaches, all have very similar applicability and differ mostly on whether the hidden message is embedded into an existing cover message or if the cover message is synthesized for it. Among the systems that use an existing cover message, Winstein's ideal coding~\cite{lex}, described in section~\ref{idealc} would have the best rate of hidden information given that it is the only system that can use all possible synonyms of words. Among systems for cover synthesis, the usage of Markov chains to generate the cover, using Huffman coding to ensure the text is statistically "natural"~\cite{markuff}, described in section~\ref{markuffsec}, is effectively the state of the art of semantic steganography  and produces the most natural results with the highest rate of hidden information.

\bibliography{bibl.bib}

\end{document}